\documentclass[twocolumn,showpacs,preprintnumbers,amsmath,amssymb]{revtex4}

\usepackage{amsmath, amssymb}
\usepackage[dvips]{graphicx}
\usepackage{bm}

\newcommand{\initialrod}{L}

\begin{document}

\title{A solvable model of fracture
with power-law distribution of fragment sizes}

\author{Ken Yamamoto}
\author{Yoshihiro Yamazaki}
\affiliation{Department of Physics, Waseda University, Tokyo, Japan}
%\date[submitted ]{\today}

\begin{abstract}
The present letter describes a stochastic model of fracture,
whose fragment size distribution can be calculated
analytically as a power-law-like distribution.
The model is basically cascade fracture,
but incorporates the effect that each fragment in each stage of cascade
ceases fracture with a certain probability.
When the probability is constant,
the exponent of the power-law cumulative distribution
lies between $-1$ and $0$,
depending not only on the probability
but the distribution of fracture points.
Whereas, when the probability depends on the size of a fragment,
the exponent is less than $-1$,
irrespective of the distribution of fracture points.
\end{abstract}


\maketitle

Fracture and fragmentation are ubiquitous in nature,
and the comprehension and control of fracture are very important
not only in science and engineering but in our daily lives.
Numerous experiments have confirmed that
fragment size distributions mainly follow
power-law distributions \cite{Oddershede, Wittel, KatsuragiIhara},
but other types, including
lognormal distributions \cite{Bindeman, Kobayashi, Andresen},
have been also observed.
From a theoretical point of view,
various models have attempted
to derive power laws \cite{Gilvarry, Cheng, Mekjian, Marsili},
but there seems to be no decisive model
which briefly and analytically explains
a power-law distribution of fragment sizes
without using specific breaking mechanisms.
Some stochastic models \cite{Sornette, Takayasu, Manrubia}
can successfully provide power-law distributions,
but they cannot be applied directly to fracture phenomena.

\begin{figure}[b!]
\flushleft{
\begin{minipage}{13.5pt}\bf(a)\vspace{2.5cm}\end{minipage}
\begin{minipage}{.4\textwidth}\flushleft
\includegraphics[scale=.65]{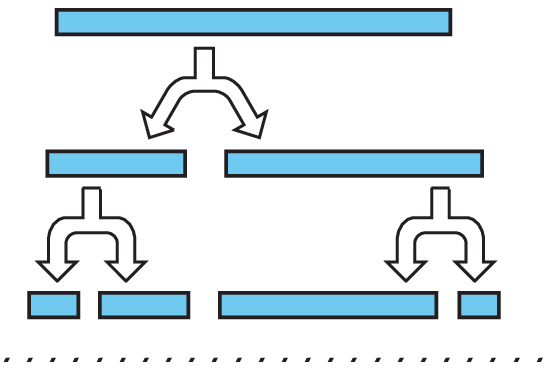}
\end{minipage}\\
\begin{minipage}{14.2pt}\bf(b)\vspace{2.5cm}\end{minipage}
\begin{minipage}{.4\textwidth}\flushleft
\includegraphics[scale=.65]{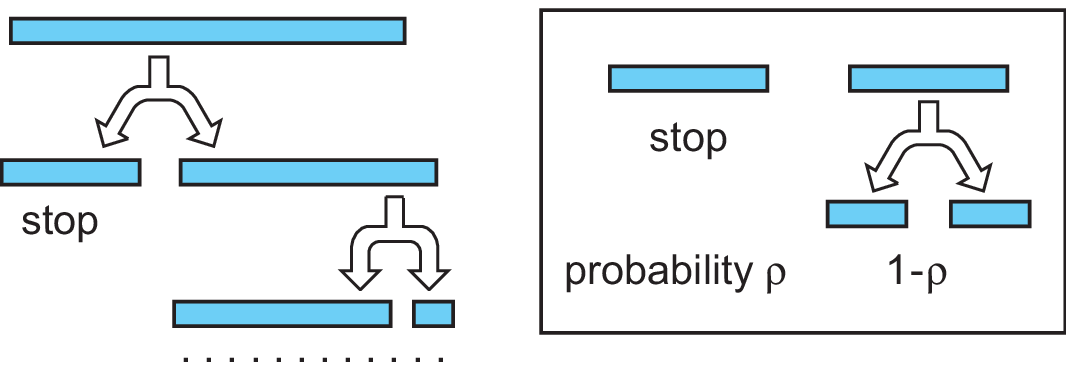}
\end{minipage}
}
\caption{
(a) The simple model of cascade fracture, where
the fragments at each stage break into two sub-fragments.
The resulting fragment size distribution is a lognormal one.
(b) Our proposed model,
which is similar to (a)
except that each fragment ceases fracture
with a constant probability $\rho$.
}
\label{fig1}
\end{figure}

% multiplicative process and lognormal distribution
A lognormal distribution of fragment sizes can be explained
by a quite simple model of cascade fracture \cite{MatsushitaSumida}.
In this model, one rod of length $\initialrod$
breaks into two fragments at a randomly chosen point,
and each of the two fragments again breaks into two sub-fragments,
and so on (see Fig. \ref{fig1} (a)).
The length of one of the fragments
after the $n$-th stage of fracture is expressed as
$\xi_1\xi_2\cdots\xi_n\initialrod$,
where $\xi_1,\xi_2,\cdots,\xi_n$ are random numbers between 0 and 1.
This process is referred to as `multiplicative',
because the length of a fragment is given
by multiplying the previous length by $\xi_i$.
Assuming that $\xi_1,\cdots,\xi_n$ are 
independently and identically distributed,
and that the variance of $\log\xi_i$ is finite,
one can prove by the central limit theorem for $\log\xi_i$
that the fragment size distribution exhibits
a lognormal distribution when $n\gg1$.

In the present letter,
we slightly modify the above multiplicative model of cascade fracture
so that
the resulting fragment size distribution becomes power-law like.

Our model also starts with one rod of length $L$,
and fragments repeatedly break into two sub-fragments.
A fracture point is given by a random number $\xi\in(0,1)$
drawn from a probability density function $g(\xi)$.
The difference from the above simple multiplicative model is that
each fragment ceases fracture with a constant probability $\rho$,
which we call the ``stopping probability'' (see Fig. \ref{fig1} (b)).
Whether each fragment stops fracture or not
is determined independently;
once a fragment ceases fracture,
it never restarts fracture any more,
and we call such a fragment ``inactive''.

In order to analyze this model,
we focus on the cumulative number $N_\initialrod(x)$ of fragments,
which represents the expected number of fragments larger than $x$.
The initial length $\initialrod$ of the rod is specified as a parameter.
$N_\initialrod(x)$ satisfies the following equation.
\begin{equation}
N_\initialrod(x)
=(1-\rho)\int_0^1\left\{N_{\xi\initialrod}(x)+N_{(1-\xi)\initialrod}(x)\right\}
	g(\xi){\mathrm d}\xi+\rho.
\label{eq:inhomogeneous}
\end{equation}
The first term (integration) at the right hand side
represents that the initial rod breaks into two fragments
of the lengths $\xi\initialrod$ and $(1-\xi)\initialrod$
with probability $(1-\rho)$,
and the last term represents that the initial rod becomes inactive
with probability $\rho$.

If we rescale our length scale by a factor $\alpha(>0)$
and observe fracture processes,
the length of the initial rod is $\alpha\initialrod$ in the new scale,
and the cumulative number $N_\initialrod(x)$ corresponds to
$N_{\alpha\initialrod}(\alpha x)$.
This seeming difference is just brought by rescaling, so
we have a scaling relation
$N_{\alpha\initialrod}(\alpha x)=N_\initialrod(x)$ or
$N_{\alpha\initialrod}(x)=N_\initialrod(x/\alpha)$.
Using this relation to convert all subscripts
in Eq. \eqref{eq:inhomogeneous} into $\initialrod$,
and introducing
\begin{equation}
\tilde{N}_\initialrod(x)=N_\initialrod(x)+\frac{\rho}{1-2\rho}
\label{eq:tilde}
\end{equation}
in order to eliminate the last term ``$\rho$''
in Eq. \eqref{eq:inhomogeneous},
we obtain
a homogeneous equation of $\tilde{N}_\initialrod$:
\[
\tilde{N}_\initialrod(x)
=(1-\rho)\int_0^1
	\left\{\tilde{N}_\initialrod\left(\frac{x}{\xi}\right)
			+\tilde{N}_\initialrod\left(\frac{x}{1-\xi}\right)
	\right\}g(\xi){\mathrm d}\xi.
\]
We assume a power-law form
$\tilde{N}_\initialrod(x)=Cx^{-\beta}$,
where $C$ and $\beta(>0)$ are both independent of $x$.
Then, we have
\begin{equation}
(1-\rho)\int_0^1\left\{\xi^\beta+(1-\xi)^\beta\right\}g(\xi){\mathrm d}\xi=1.
\label{eq:beta}
\end{equation}
The exponent $\beta$ is determined by this equation;
hence $\beta$ generally depends on both $\rho$ and $g$.
Furthermore,
it is natural to assume that
the fracture is left-right symmetrical if a rod is uniform,
that is, $g(\xi)=g(1-\xi)$ for any $\xi\in(0,1)$.
This condition simplifies Eq. \eqref{eq:beta} to
\begin{equation}
2(1-\rho)\int_0^1 \xi^\beta g(\xi){\mathrm d}\xi=1.
\label{eq:beta_simple}
\end{equation}

% the derivation of the coefficient C
We finally determine the coefficient $C$.
By the definition of the cumulative number,
$N_\initialrod(\initialrod)$ is the expected number of fragments
larger than $\initialrod$,
and it is equal to the probability
with which the initial rod ceases fracture.
Thus, $N_\initialrod(\initialrod)=\rho$.
On the other hand,
$\tilde{N}_\initialrod(\initialrod)=C\initialrod^{-\beta}$
is immediately obtained.
Therefore, it follows by Eq. \eqref{eq:tilde} that
\[
C=\frac{2\rho(1-\rho)}{1-2\rho}\initialrod^\beta.
\]
Eventually, the complete solution is
\begin{align*}
N_\initialrod(x)&=\frac{2\rho(1-\rho)}{1-2\rho}\initialrod^\beta x^{-\beta}
	-\frac{\rho}{1-2\rho}\\
&=\frac{\rho}{1-2\rho}
	\left\{2(1-\rho)\left(\frac{x}{\initialrod}\right)^{-\beta}-1\right\},
\end{align*}
coupled with Eq. \eqref{eq:beta} for the determination of $\beta$.
Note that $\tilde{N}_\initialrod$ is an exact power law,
but $N_\initialrod$ is not exactly
because of the presence of the second term ``$-1$''.
Nonetheless,
$N_\initialrod(x)$ can be approximated by a power law
if the second term is negligible,
i.e., $2(1-\rho)\gg(x/\initialrod)^\beta$,
or if $x$ is sufficiently smaller than $\initialrod$
and $\rho$ is also small.

The solution $\beta$ of Eq. \eqref{eq:beta} or \eqref{eq:beta_simple}
cannot be expressed explicitly
for general probability density $g$.
We provide two examples of calculations of $\beta$.
The mathematically simplest instance is
$g(\xi)=\delta(\xi-1/2)$, where $\delta$ is the Dirac delta function.
In other words,
the fracture points are at the middle of the fragments.
Equations \eqref{eq:beta} and \eqref{eq:beta_simple} in this case
are both reduced to
\[
2(1-\rho)\cdot \left(\frac{1}{2}\right)^\beta=1,
\]
and the solution is $\beta=1+\log(1-\rho)/\log2$.

In the second example,
the fracture point is distributed uniformly over each fragment,
i.e., $g(\xi)=1$ for all $\xi\in(0,1)$.
Equation \eqref{eq:beta_simple} becomes
\[
2(1-\rho)\int_0^1\xi^\beta{\mathrm d}\xi
=2(1-\rho)\frac{1}{1+\beta}=1,
\]
and the solution is $\beta=1-2\rho$.

Both in these two specific examples,
$\beta=1+\log(1-\rho)/\log2$ and $\beta=1-2\rho$,
the possible ranges of $\beta$ and $\rho$ are restricted by each other:
$\rho<1/2$ follows from $\beta>0$,
and $\beta<1$ follows from $\rho>0$.
Namely,
the reasonable value of the stopping probability is $0<\rho<1/2$
and the possible value of $\beta$ is within $0<\beta<1$.

\begin{figure}[h!]\centering
\includegraphics[width=.3\textwidth]{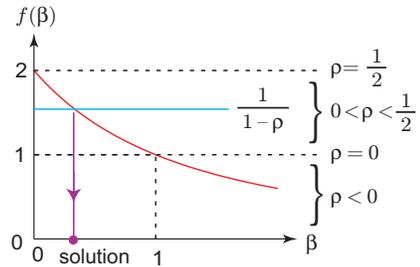}
\caption{
The structure of Eq. \eqref{eq:beta}.
The unique solution $\beta\in(0,1)$
exists for any $\rho\in(0,1/2)$.
}
\label{fig2}
\end{figure}

The same restrictions for $\rho$ and $\beta$
also hold for a general probability density $g$.
For a fixed $g$, we consider
$f(\beta)
	:=\int_0^1\left\{\xi^\beta+(1-\xi)^\beta\right\}g(\xi){\mathrm d}\xi$,
which is defined at least in $\beta\ge0$, and continuous and differentiable.
Clearly, Eq. \eqref{eq:beta} is expressed as $f(\beta)=1/(1-\rho)$.
Using the normalization $\int_0^1 g(\xi){\mathrm d}\xi=1$,
we have $f(0)=2$ and $f(1)=1$ for any $g$.
Hence, the intermediate value theorem in elementary calculus
insures that the equation
$f(\beta)=1/(1-\rho)$ has {\it at least} one solution $\beta\in(0,1)$
if $1<1/(1-\rho)<2$ (or $0<\rho<1/2$).
On the other hand,
$f(\beta)$ is decreasing because
\[
\frac{\mathrm d}{{\mathrm d}\beta}f(\beta)
=\int_0^1\{\xi^\beta\log\xi+(1-\xi)^\beta\log(1-\xi)\}g(\xi){\mathrm d}\xi<0,
\]
therefore the correspondence
between $\beta$ and $\rho$ is one-to-one,
i.e., $\beta$ can be determined {\it uniquely} for given $\rho$.
Positivity $\rho>0$ holds in $\beta<1$,
and $\beta>0$ holds in $1/(1-\rho)<2$ (or $\rho<1/2$).
See Fig. \ref{fig2} for the reference of the analysis.

\begin{figure}[]
\centering
\begin{minipage}{2em}{\flushleft\bf(a)\vspace{3.5cm}}\end{minipage}
\begin{minipage}{.3\textwidth}
\includegraphics[width=\textwidth]{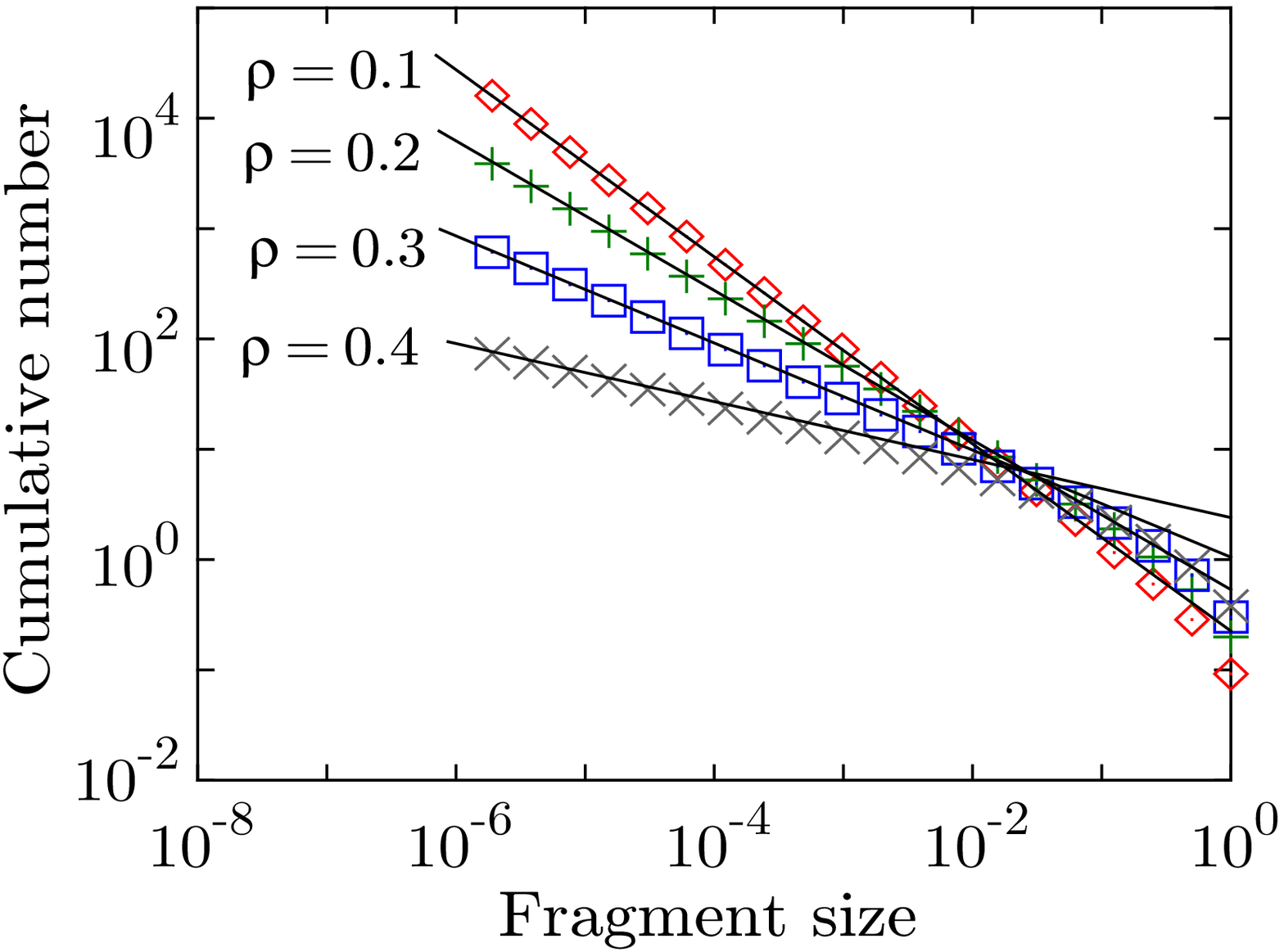}
\end{minipage}\\
\vspace{\baselineskip}
\begin{minipage}{2em}{\flushleft\bf(b)\vspace{3.5cm}}\end{minipage}
\begin{minipage}{.3\textwidth}
\includegraphics[width=\textwidth]{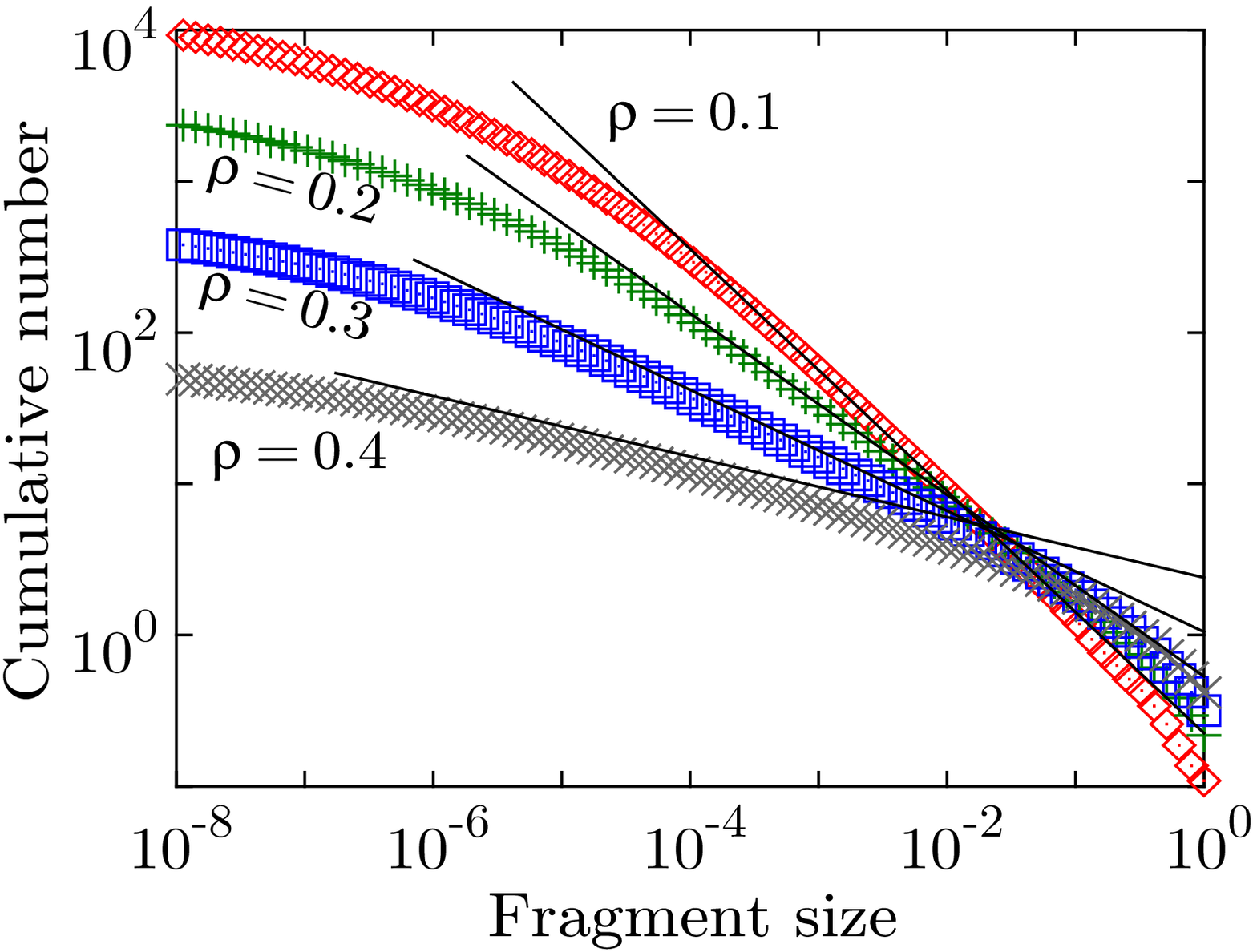}
\end{minipage}
\caption{
Numerical result of cumulative number $N_\initialrod(x)$
for $\initialrod=1$, and $\rho=0.1,0.2,0.3,$ and 0.4,
generated by counting only the inactive fragments within 20th stage of cascade,
and averaging 1000 samples each.
Each straight line indicates the corresponding $\tilde{N}_\initialrod$
that follows an exact power law.
The fracture points are 
(a) at the middle of the fragments $g(\xi)=\delta(x-1/2)$,
and (b) distributed uniformly $g(\xi)=1$.
}
\label{fig3}
\end{figure}

\begin{figure*}\centering
\begin{minipage}{2em}{\flushleft\bf(a)\vspace{4cm}}\end{minipage}
\begin{minipage}{.3\hsize}
\includegraphics[width=\textwidth]{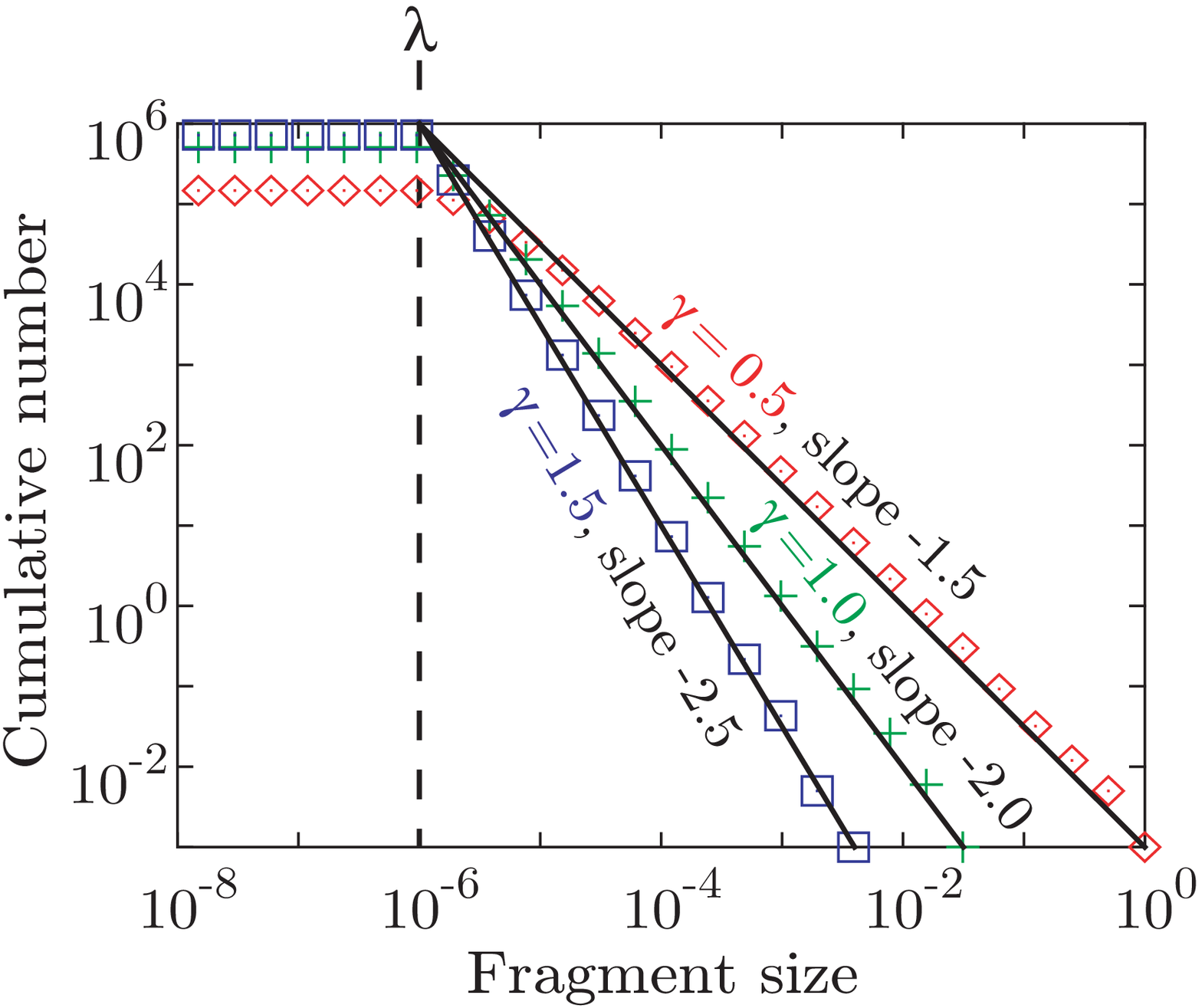}
\end{minipage}
\hspace{.05\textwidth}
\begin{minipage}{2em}{\flushleft\bf(b)\vspace{4cm}}\end{minipage}
\begin{minipage}{.3\hsize}
\includegraphics[width=\textwidth]{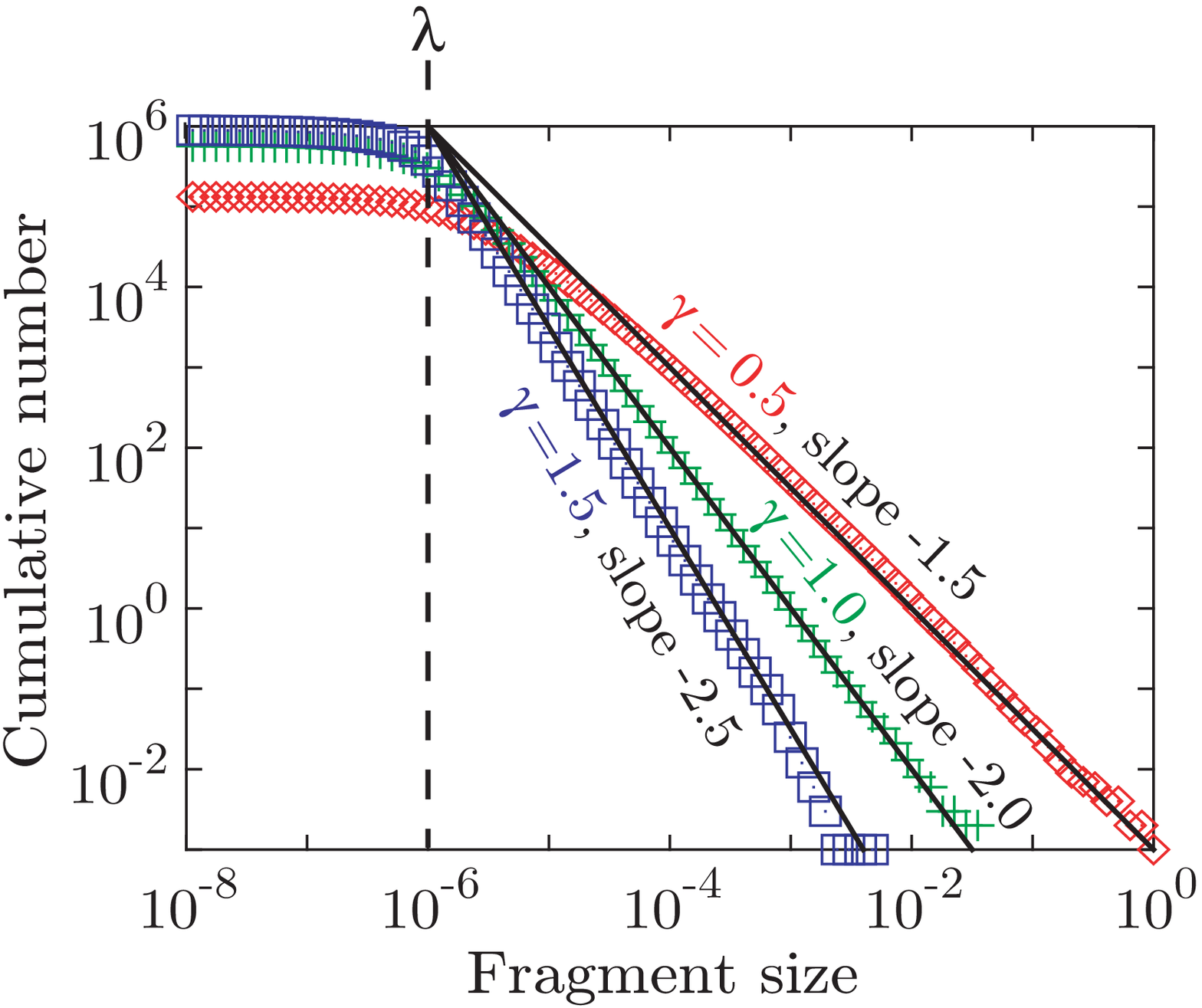}
\end{minipage}
\caption{
Numerical results of $N_\initialrod$ for $\initialrod=1$.
The parameter values of the stopping probability are
$\lambda=10^{-6}$ and $\gamma=0.5,1,$ and $1.5$.
Each data set was generated by averaging 1000 samples.
Solid lines indicate the corresponding solutions \eqref{eq:solution}.
The probability densities for the fracture points are respectively
$g(\xi)=\delta(\xi-1/2)$ in (a), and $g(\xi)=1$ in (b).
}
\label{fig4}
\end{figure*}

Figure \ref{fig3} shows numerical results of $N_\initialrod$.
The parameters are $\initialrod=1$, and $\rho=0.1,0.2,0.3,0.4$.
The probability density $g$ for the fracture points
are $g(\xi)=\delta(\xi-1/2)$ in (a) and $g(\xi)=1$ in (b).
We limited fracture to 20 stages at the maximum,
and we counted only the inactive fragments.
Each plot is the average of 1000 samples.
An exact power law $\tilde{N}_\initialrod$ is
also shown with black lines.
Power laws fail in larger fragment sizes, as mentioned above.
In the panel (b),
the cumulative numbers also deviate largely from power laws in smaller sizes
because the number of fracture steps is bounded:
some long fragments are still active
after the last fracture stage in the simulation,
and they will contribute to raising of the number of small fragments
if fracture is continued.

It is noted that there have been many experimental results of $\beta>1$,
but the above model provides only $\beta<1$.
Here we modify the model in order to realize $\beta>1$.
Recalling the above analysis,
we have treated the stopping probability as a constant value $\rho$.
Instead, we set here
the stopping probability as a function of a fragment size.
In particular, we give
the stopping probability of a fragment of size $\ell$ as
\begin{equation}
p_\lambda(\ell)=
\begin{cases}
\left(\frac{\lambda}{\ell}\right)^\gamma, & (\ell\le\lambda)\\
1, & (\ell\ge\lambda)\\
\end{cases}
\label{eq:p_lambda}
\end{equation}
where $\lambda$ is a characteristic length
and $\gamma>0$ is a constant.
It represents an effect that
smaller fragments are more difficult to experience further fracture.
Obviously,
a fragment becomes inactive whenever its size becomes smaller than $\lambda$,
hence the parameter $\lambda$ is the lower bound of the fragment sizes.
We employ the assumption $\lambda\ll\initialrod$ in the following analysis.

As above,
the cumulative number 
$N_{\initialrod,\lambda}(x)$,
including two parameters $\initialrod$ and $\lambda$ this time,
plays an important role in the following analysis.
In the same way as Eq. \eqref{eq:inhomogeneous},
$N_{\initialrod,\lambda}$ satisfies the following equation.
\begin{align}
&N_{\initialrod,\lambda}(x)\nonumber\\
&=\left\{1-\left(\frac{\lambda}{\initialrod}\right)^\gamma\right\}
	\int_0^1\left\{N_{\xi\initialrod,\lambda}(x)
					+ N_{(1-\xi)\initialrod,\lambda}(x)\right\}
			g(\xi){\mathrm d}\xi\nonumber\\
&\qquad+ \left(\frac{\lambda}{\initialrod}\right)^\gamma\nonumber\\
&\simeq\int_0^1\left\{N_{\xi\initialrod,\lambda}(x)
					+ N_{(1-\xi)\initialrod,\lambda}(x)\right\}
			g(\xi){\mathrm d}\xi,
\label{eq:homogeneous}
\end{align}
where we used the approximation $\lambda/\initialrod\simeq0$.
(the symbol ``$\simeq$'' is used only in this sense.)

A scaling relation
$N_{\alpha\initialrod,\alpha\lambda}(\alpha x)=N_{\initialrod,\lambda}(x)$
is again obtained.
We need another scaling relation for the analysis.
By the definition of the cumulative number,
\begin{widetext}
\begin{align*}
N_{\initialrod,\lambda}(x)
&=\sum_{n=0}^\infty P(\xi_1\cdots\xi_n\initialrod\ge x)
	\{1-p_\lambda(\initialrod)\}\{1-p_\lambda(\xi_1\initialrod)\}
	\cdots\{1-p_\lambda(\xi_1\cdots\xi_{n-1}\initialrod)\}
	\cdot p_\lambda(\xi_1\cdots\xi_n\initialrod)\\
&\simeq\sum_{n=0}^\infty P(\xi_1\cdots\xi_n\initialrod\ge x)
	\left(\frac{\lambda}{\xi_1\cdots\xi_n\initialrod}\right)^\gamma
\propto \lambda^\gamma.
\end{align*}
\end{widetext}
Thus,
$N_{\initialrod,\alpha\lambda}(x)
	\simeq\alpha^\gamma N_{\initialrod,\lambda}(x)$
is derived for $x\gg\lambda$ and $\alpha>0$.
We guess 
a power-law form $N_{\initialrod,\lambda}(x)=Cx^{-\beta}$,
and substitute into Eq. \eqref{eq:homogeneous}
together with two scaling relations,
which yields
\[
\int_0^1\left\{\xi^{\beta-\gamma}+(1-\xi)^{\beta-\gamma}\right\}g(\xi)
	{\mathrm d}\xi=1.
\]
This equation can be solved immediately as $\beta=1+\gamma$,
where we note the normalization $\int_0^1 g(\xi){\mathrm d}\xi=1$.
$\beta>1$ is attained because $\gamma>0$.
A remarkable point is that
the exponent $\beta=1+\gamma$ is universal
over any probability density $g$ governing the fracture points.
(Compare with the case of a constant stopping probability,
where $\beta$ depends on $g$.)

% The determination of C should be placed after \beta is obtained
The coefficient $C$ is $\lambda^\gamma\initialrod$,
derived from the consistency of two expressions
$N_{\initialrod}(\initialrod)=C\initialrod^{-\beta}=CL^{-(1+\gamma)}$ and
$N_{\initialrod}(\initialrod)
	=p_\lambda(\initialrod)=(\lambda/\initialrod)^\gamma$.
Finally, the complete solution is expressed as
\begin{equation}
N_{\initialrod,\lambda}(x)
=\lambda^\gamma\initialrod x^{-(1+\gamma)}
=\left(\frac{\lambda}{\initialrod}\right)^\gamma
	\left(\frac{x}{\initialrod}\right)^{-(1+\gamma)}.
\label{eq:solution}
\end{equation}
The calculation is based on $x\gg\lambda$;
consequently,
this solution probably breaks down if $x\lesssim\lambda$.

Numerical results are shown in Fig. \ref{fig4},
where we set $\initialrod=1$ and $\lambda=10^{-6}$.
Numerically-generated cumulative numbers
clearly lie on the power-law solution (solid line)
over a wide range of larger fragment sizes.
Also,
the data points deviate from the power laws
in a fragment size close to or less than $\lambda$,
as expected theoretically.

% Discussion
One can straightforwardly extend the model
so that each fragment breaks into $n$ subfragments at a single fracture,
where $n$ can be either a fixed or random number.
A fragment size distribution in this case is again like a power law;
the exponent $\beta$ is less than 1 under a constant stopping probability,
and $\beta=1+\gamma(>1)$ under the stopping probability
as in Eq. \eqref{eq:p_lambda}.
A special case like the Sierpinski fractal
is found in Refs. \cite{Matsushita, Kadono}
without pointing out the sensitivity of $\beta$ against $g$.

Our model claims that a lognormal and power-law distribution are similar;
the difference is whether the stopping probability exists or not.
Their similarity has been supported experimentally.
A fragment size distribution qualitatively changes
according to impact energy \cite{Katsuragi}
(or falling height \cite{Ishii}):
it exhibits a lognormal distribution under lower energy,
and a power-law distribution under higher energy.
These results imply that
a lognormal distribution and a power-law distribution
can possess a common origin.
Furthermore, the proposed mechanism,
where a multiplicative stochastic process with random stopping
produces a power-law distribution, is quite simple and general,
so it will be applicable to other systems than fracture.

The present work was supported by
Grant-in-Aid for JSPS Fellows from
the Japan Society for the Promotion of Science.
We are grateful to Dr. Mitsugu Matsushita and Dr. Naoki Kobayashi
for their informative discussions.


\begin{thebibliography}{10}
\bibitem{Oddershede}
L. Oddershede, P. Dimon, and J. Bohr, Phys. Rev. Lett. 71, 3107 (1993).
\bibitem{Wittel}
F. Wittel, F. Kun, H.J. Herrmann, and B.H. Kr\"oplin,
Phys. Rev. Lett. 93, 035504 (2004).
\bibitem{KatsuragiIhara}
H. Katsuragi, H. Honjo, and S. Ihara, Phys. Rev. Lett. 95, 095503 (2005).
\bibitem{Bindeman}
I.N. Bindeman, American Mineralogist 90, 1801 (2005).
\bibitem{Kobayashi}
N. Kobayashi, K. Kohyama, Y. Sasaki, and M. Matsushita,
J. Phys. Soc. Jpn. 75, 083001 (2006).
\bibitem{Andresen}
C. A. Andresen, A. Hansen, and J. Schmittbuhl, Phys. Rev. E 76, 026108 (2007).
\bibitem{Gilvarry}
J.J. Gilvarry and B.H. Bergstrom, J. Appl. Phys. 32, 400 (1961).
\bibitem{Cheng}
Z. Cheng and S. Redner, Phys. Rev. Lett. 60, 2450 (1988).
\bibitem{Mekjian}
A.Z. Mekjian, Phys. Rev. Lett. 64. 2125 (1990).
\bibitem{Marsili}
M. Marsili and Y.-C. Zhang, Phys. Rev. Lett. 77, 3577 (1996).
\bibitem{Sornette}
D. Sornette and R. Cont, Journal de Physique I 7, 431 (1997).
\bibitem{Takayasu}
H. Takayasu, A.-H. Sato, and M. Takayasu, Phys. Rev. Lett. 79, 966 (1997).
\bibitem{Manrubia}
S.C. Manrubia and D.H. Zanette, Phys. Rev. E 59, 4945 (1999).
\bibitem{MatsushitaSumida}
M. Matsushita and K. Sumida, Bull. Facul. Sci. Eng. Chuo Univ.
31, 69 (1988).
\bibitem{Matsushita}
M. Matsushita, J. Phys. Soc. Jpn. 54, 857 (1985).
\bibitem{Kadono}
T. Kadono and M. Arakawa, Phys. Rev. E 65, 035107(R) (2002).
\bibitem{Katsuragi}
H. Katsuragi, D. Sugino, and H. Honjo, Phys. Rev. E 70, 065103(R) (2004).
\bibitem{Ishii}
T. Ishii and M. Matsushita, J. Phys. Soc. Jpn. 61, 3474 (1992).
\end{thebibliography}
\end{document}